\documentstyle[prl,aps,multicol,epsf]{revtex}

\begin{document}
\draft
\tighten

\title{Geometry and topological order in the non-relativistic 
Luttinger liquid.}
\author{H.V. Kruis, I.P. McCulloch, Z. Nussinov and J. Zaanen}
\address{Instituut-Lorentz for Theoretical Physics, Leiden University\\
P.O.B. 9506, 2300 RA Leiden, The Netherlands}
\date{\today ; E-mail:hvkruis@lorentz.leidenuniv.nl;
zohar@lorentz.leidenuniv.nl; jan@lorentz.leidenuniv.nl
}
\maketitle

\begin{abstract}
We demonstrate that the notion of spin-charge separation as follows
from the Tomonaga-Luttinger theory of relativistic fermions does not
suffice to completely specify the Luttinger liquid state associated 
with the fermions of condensed matter physics. The latter carries in
addition a form of topological order which can be made
visible using non-local correlation functions familiar 
from lattice gauge theory.  
 \end{abstract}

\begin{multicols}{2}
\narrowtext

\setlength{\textfloatsep}{10cm}

Systems composed of an infinite number of interacting quantum particles can
be characterized by long wavelength excitations carrying quantum
numbers unrelated to those of the constituents. A classic example 
is the phenomenon of spin-charge separation which is well established 
in one dimensional electron systems\cite{bosoref}. 
The elementary excitations can
be viewed as pieces of the original electron, carrying separately the
charge and the spin. This notion first emerged when 
it was demonstrated that
systems of relativistic fermions in 1+1D can be represented exactly
under a variety circumstances in terms of Bose fields,
carrying separately the spin and the charge degrees of freedom 
(bosonization). Although a-priori it is unclear whether the relativistic 
theory has anything to say about systems of non-relativistic
electrons, an overwhelming case emerged that bosonization is
correctly describing the asymptotic properties of the local
correlation functions of the latter. However, the loophole
is that these might  differ by  a non-critical  topological 
order which cannot be detected using two-point correlators in terms of
local operators. Here we will demonstrate that at least the Hubbard model
in 1+1D carries such an order which is alien to relativistic fermions. 
We believe that the Hubbard model is in this regard fully representative
for all non-relativistic electron systems.

This hidden order has to do with the fact that the charge carriers
(`holons') are at the same time  Ising `domain walls' (kinks) in the
spin system, a property shared with stripes in two dimensional
systems\cite{philmag}. Resting on the
{\em geometrical} definition of spin-charge separation 
as deduced from the Bethe-Ansatz\cite{liebwu} solution of the Hubbard model
at infinite $U$, obtained first by
Woynarovich\cite{woyn} and later Ogata and Shiba\cite{ogashi}, 
we construct a non-local operator measuring directly
the topological order. Using a state of the art DMRG method, we demonstrate
numerically that this order is present in the Hubbard chain for all
positive $U$'s and electron densities. We subsequently show that  a
straightforward application of bosonization leads to inconsistencies,
demonstrating that this topological order is not present in the Dirac
vacuum in 1+1D. Our non-local correlator has the structure of the Wilson
loop of a $Z_2$ gauge theory and we conclude with the speculation that 
a specific kind of 1+1D superconductor might be governed by a local
Ising symmetry.
 
Some time ago\cite{woyn,ogashi}, a simple but 
most peculiar  property of the Bethe-Ansatz solution of the 1D Hubbard
model in the limit that $U \rightarrow \infty$ was found; namely that 
the wave function factorizes into a charge part $\psi_{SF}$, depending 
only on where the electrons are, and a spin part $\psi_{H}$
which merely depends on the way the spins are distributed,
\begin{equation}
\psi(\{ x_{i} \}_{i=1}^{N}; \{ y_{j} \}_{j=1}^{N/2}) = \psi_{SF}
(\{x_{i} \}_{i=1}^{N})~ \psi_H (\{ y_{j} \}_{j=1}^{N/2}).
\label{ogatashiba}
\end{equation}
The charge part $\psi_{SF}$ is the wave-function of a
non-interacting spinless-fermion system where the coordinates $x_i$ refer
to the positions of the electrons/spinless fermions. The spin part
$\psi_H$ is identical to the wavefunction of a chain of Heisenberg spins
interacting via a nearest-neighbor antiferromagnetic exchange. 
In $\psi_H$ only the positions of the up spins are needed and these 
correspond with the coordinates $y_i$. However, the miracle is 
that the coordinates $y_i$ do not refer to the 
Hubbard chain, but instead to a {\em new space}: a
lattice with sites at coordinates $x_{1},  x_{2}, ..., x_{N}$ given by  
the positions of the charges in a configuration with 
amplitude $\psi_{SF}$. 

This shows that the
quantum dynamics of interacting electrons generates a {\em geometrical 
structure} analogous to the fabric of general relativity.
Let us visualize this for a representative example (Fig. 1). Consider
 $N$ electrons
on a chain with $L$ sites under the condition that $N < L$ such that the charge
configurations can be specified by the locations of the holes.
A charge configuration in the full Hubbard chain (`external space'), 
has an amplitude  $\psi_{SF}$ in the wavefunction with the 
coordinates of the dots corresponding 
with the $x_i$'s. The spin system sees a different `internal space' 
obtained from the full space  by removing the holes {\em together with the
sites where the holes are located}, substituting the hole and its site with an
antiferromagnetic exchange between the sites neighboring the hole (the
`squeezed space'\cite{ogashi}).  

\begin{figure}[b]
\hspace{0.0 \hsize}
\epsfxsize=0.9\hsize
\epsffile{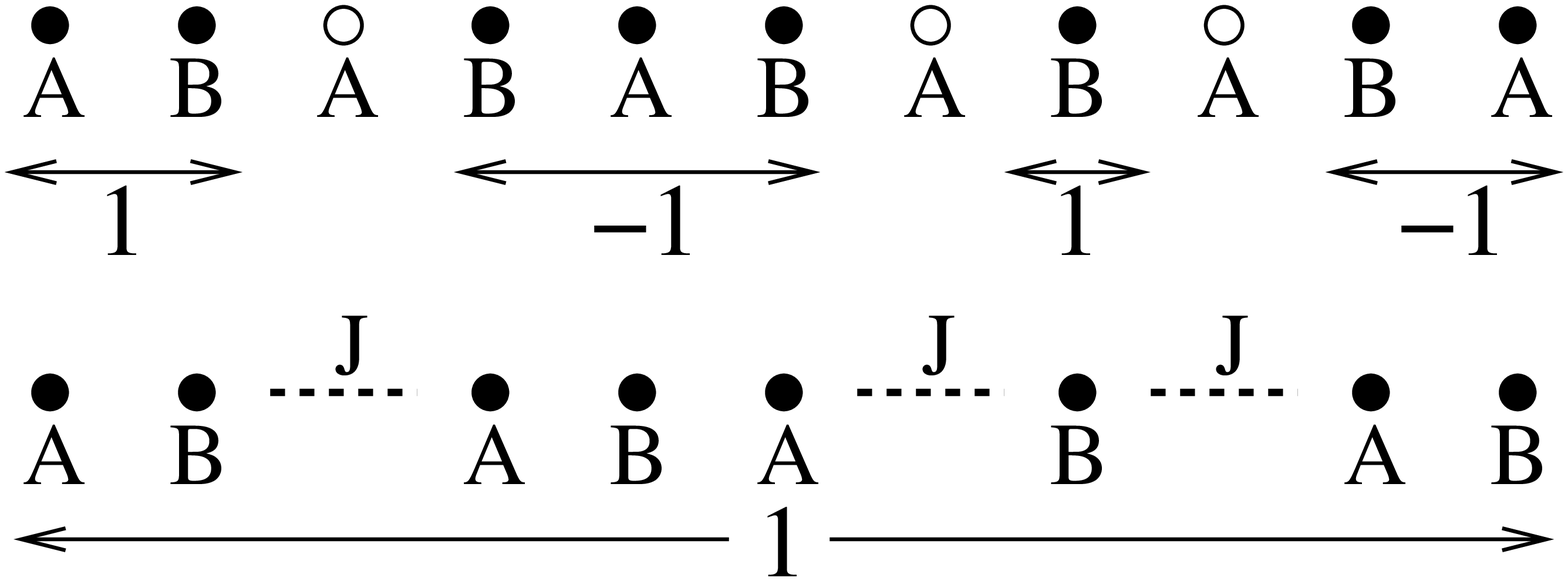}
\caption{The emergence of the $Z_2$ `sublattice parity field' in the
geometrical `squeezing' operation. The open circles refer to some
configuration of holes. The black dots define the embedding space of
the spin system in both the full- (upper) and squeezed (lower) lattice.
The unsqueezing operation can be parametrized in terms of binding of
the electric charge quantum to flips in the $Z_2$ valued sublattice
parity field.} 
\label{f1}
\end{figure}

This is a very simple geometrical structure which can be formulated
in terms of a simple topological `gauge' theory. 
In what regards are the full chain and the squeezed chain different? 
The squeezed chain is obviously shorter than the full chain and this is
a simple dilation: a distance $x$ measured in the full chain becomes a distance
$\rho x$ in the squeezed chain ($\rho = N / L$, the electron density)
 when $x \gg 1$,
the lattice constant. The other aspect is also simple, but less trivial in its
consequences. The spin system is a quantum-antiferromagnet and is therefore
sensitive to the geometrical property of bipartiteness. 
 A lattice is called bipartite if it can be subdivided 
in two sublattices $A$ and $B$, such that all sites on the $A$ sublattice
are neighbored by $B$ sublattice sites and vice versa. This division can be
done in two ways ($\cdots - A - B - A - B \cdots$ and $\cdots - B - A - B - A
\cdots$) defining a $Z_2$ valued quantity `sublattice parity', $p = \pm 1$. 
Consider now  what happens with  sublattice parity in the squeezing. For the
Heisenberg spin chain a redefinition of $p = 1
\leftrightarrow -1$ does not carry any consequence (`pure gauge').
However, sublattice parity becomes important
in the mapping of squeezed space into full space (Fig. 1). `Fix the gauge'
in squeezed space by choosing a particular sublattice parity, and consider
what happens when it is unsqueezed. The holes are inserted, and because every  
hole is attached to one site,  every time a hole is passed the
sublattice parity flips.  
This is true for every instantaneous charge configuration, but the ground
state is a  a superposition of many of these configurations: charge is 
fluctuating and since it is attached to the sublattice parity flips, 
the full space which is observable
by external observers (experimentalists) should be considered as a fluctuating 
geometry. However, this is a very simple fluctuating geometry 
because all that is fluctuating is the property of bipartiteness. 

Let us now ask the question if a correlation
function can be defined acting on the full Hubbard chain which can measure
the true spin correlations living in squeezed space. Since all that
matters is sublattice parity this can be achieved by simply multiplying
the spin operator by a factor $-1$ every time a hole is passed, thereby 
removing the sublattice parity flips from the spin correlations. Define
staggered magnetization as $\vec{M} (x_i) = (-1)^{x_i} \vec{S} (x_i)$ where
$\vec{S}$ is the spin operator ($S^z = \frac{1}{2}(n_{\uparrow} - n_{\downarrow}),
S^{+} = c^{\dagger}_{\uparrow} c_{\downarrow}$) with the charge operator 
$n_{x_i} = n_{x_i \uparrow} + n_{x_i \downarrow}$ taking the values 0,
1 and 2 for an empty-, singly- and doubly occupied sites respectively. The
correlation function we are looking for is \cite{zavsa},
\begin{eqnarray}
O_{top} (x) = \langle M^{z} (0)
(-1)^{\sum_{j=1}^{x-1} 1 - n_j } M^{z} (x)  \rangle.
\label{otop}
\end{eqnarray}
The operator $(-1)^{( 1 - n_j)}$ takes the value $+1$ for a singly
occupied site while it is $-1$ for a charge (hole, or
doubly occupied) site. By multiplying these values on the interval 
$0 < j <  x$ all the minus signs associated with the sublattice
parity flips are removed from the spin correlations.

Although the `string' operator $(-1)^{\sum_j 1 - n_j}$
is non-local
it can be evaluated in a straightforward manner using the techniques introduced by
Parola and Sorella\cite{parsor}. It is easily shown that in the large
$U$ limit,
\begin{eqnarray}
\langle M^{z}(0)\, &(-1)&^{\sum_{j=1}^{x-1} 1- n_j } 
\, M^{z}(x) \, \rangle 
\nonumber \\ 
&=& \sum_{j=2}^{x+1} P_{SF}^{x} (j)
(-1)^{j} O_{Heis.} (j-1)
\label{parsor}
\end{eqnarray}
where $O_{Heis.}$ is the spin correlator of the Heisenberg chain,
while $P^{x}_{SF} (j) = \langle \,  n(0) \,  n(x) \,
\delta (\sum_{l=0}^{x} n_l - j) \, \rangle _{SF}$ is the probability of
 finding $j$ spinless fermions in the interval $\left[0,x \right]$. This
factor causes the additional decay of the spin correlations due to the
charge fluctuations in the standard spin correlator. However, it is easily 
shown that it is precisely compensated in $O_{top}$ by the factor 
$(-1)^{j}$ coming from the string operator and we find the result, 
asymptotically exact for large distances $x$,
\begin{equation}
O_{top} ( x ) = {\rho \over x } \ln^{1/2} ( \rho x ).
\label{hubtop}
\end{equation}
Which is identical to the result for the pure Heisenberg spin chain
${1 \over x} \ln^{1/2} (x)$ after rescaling the amplitude of staggered
spin $\vec{M} \rightarrow {{\vec{M}} \over {\rho}}$ and the measure of length
$x  \rightarrow x / \rho$, where $\rho$ is the average charge density. 
Comparing Eq.(\ref{hubtop}) with the well-known\cite{parsor} asymptotic behavior of the 
local staggered spin correlations for the Hubbard model, 
$ O_{\vec{M}} (x)   =  \langle | M^{z} (0) M^{z}(x) | \rangle
\sim { { \cos ( [ 2 k_F - \pi] x) } \over x^{1 + K_{\rho}} }$
we find that the former decays more slowly.
 The charge-stiffness $K_{\rho}$ is associated 
with the decay of the charge correlations, $\langle n (0) n (x) \rangle
\sim  \cos ( [4 k_F - \pi] x)  / x^{4K_\rho}$, and $K_{\rho} > 0$ 
for all $\rho \neq 1$. Hence, the charge 
fluctuations modify the spin correlations because the charge is attached 
to the sublattice parity flips, resulting in a simple 
multiplicative factor $1 / x^{K_{\rho}}$. 

Is the above an accident of the strongly coupled case? Having identified
the operator Eq. (\ref{otop}) which in combination with the local 
spin-spin correlator  measures directly the presence or absence of the
topological order it becomes possible to study it in any one dimensional
system. For finite $U$'s it becomes very hard to calculate Eq. (\ref{otop})
from the Bethe-Ansatz solution and therefore we compute it numerically
for the Hubbard model using the density matrix renormalization
group. We employ an algorithm recently developed by one of us\cite{DMRG}, 
making
explicit use of the full non-abelian $SO(4)$ symmetry of the Hubbard model. In
this formulation, the generators of the global symmetry group are 
the spin $\vec{S}$ and the pseudospin $\vec{I}$; the latter is a
generalization from the charge $U(1)$ to an $SU(2)$ symmetry such
that the particle number at site $i$ is given by $n_i = 1 + 2 I^z_i$.

In principle, it is straightforward to calculate the expectation value of
string operators such as Eq. (\ref{otop}) using
the DMRG method. However, the calculation of critical correlators
with DMRG is not unproblematic even in 1+1D, because eventually the
truncation errors inherent in the form of the wavefunction will
cause an exponential decay\cite{RommerOstlund}, and the requirement
to use open boundaries causes a loss of translational invariance
even at large distances. We used a relatively large
system size (1000 sites) to reduce the effect of the boundary conditions,
and used a basis size (700 $SO(4)$ states) large enough to 
achieve a truncation length of the order of 200-300 lattice constants.
In this way meaningful results can be obtained from a simple curve fit to obtain
the desired exponents. 

In  Fig. 2 we show our results for the exponents of both the 
standard two point spin correlator $\langle \vec{M} (0) \vec{M} (x) \rangle$,
and our topological correlator, Eq. (\ref{otop}), as a function of
density for various
interaction strengths. The former illustrates 
the quality of the calculation; the dominant  correlation is at
the wavevector $2k_F - \pi$, $\sim \cos ([2k_F - \pi] x) / x^{\eta}$ and it is
seen from Fig. 2 that the exponent $\eta$ depends strongly on
the parameters. In fact, the exponent behaves exactly according
to the expectations: $\eta = K_{\sigma} + K_{\rho}$ where 
$K_{\sigma} = 1$ and $K_{\rho}$ is consistent with the values 
previously obtained by Schulz\cite{Schulz} for the same
values of $U$ and $\rho$. Considering now the topological
correlator, the dominant component lives at the
wavevector $q=0$; we find no other characteristic momenta in the Fourier
transform of Eq. (\ref{otop}) which is already reminiscent
of the staggered spin correlator of a Heisenberg chain. 
This is further
amplified by our finding that the exponent $\eta_{top}$ {\em does not depend
on the microscopic parameters at all}. In fact, for all
parameters the exponent of the topological 
correlator $\eta_{top} = 1 = K_{\sigma}$. Hence, regardless
the values of $U$ and $\rho$ the long distance behavior
of the topological correlator is indistinguishable from
the spin-spin correlator of a Heisenberg chain, demonstrating
that the Hubbard model indeed carries the sublattice parity
topological order fully for all values $U > 0$.

Let us now explain the reasons why bosonization fails in principle to
describe the topological order. Tomonaga-Luttinger
bosons cannot carry topological order for the elementary reason that
these parametrize a relativistic theory. 
At the
center of our considerations for the lattice model is the fact that
the topological structure is $Z_2$ valued (the sublattice parity) and
this in turns rests on the fact that on a lattice the number operator
$n_i$ is integer valued: a site is either empty, singly or doubly
occupied corresponding with charge quantum numbers 0,1, and 2; by
exponentiation one takes out the evenness or unevenness by the
factors $\sigma^3_i = (-1)^{n_i} = \pm 1$. In a relativistic theory
one has to keep the fermion creation- and annihilation operator
apart, $n_r \rightarrow 
\psi^{\dagger} (r) \psi (r + \epsilon)$ and the resulting Schwinger
terms are the building blocks for bosonization.  
Thereby, the charge density becomes the gradient
of a continuous function $n ( x) = (1 / \pi) \partial_x$. Accordingly,
$(-1)^{n_i} \sim (-1)^{\partial_x \phi}$ is 
a complex scalar and not an Ising valued operator. This
reflects the fact it is impossible to
localize an electron in space in a relativistic theory, while the ability 
to ascribe a position to the electron is at the heart of the 
squeezing procedure on the lattice.

\begin{figure}[tb]
\begin{center}
\vspace{10mm}
\leavevmode
\epsfxsize=0.9\hsize
\epsfysize=10cm
\epsfbox{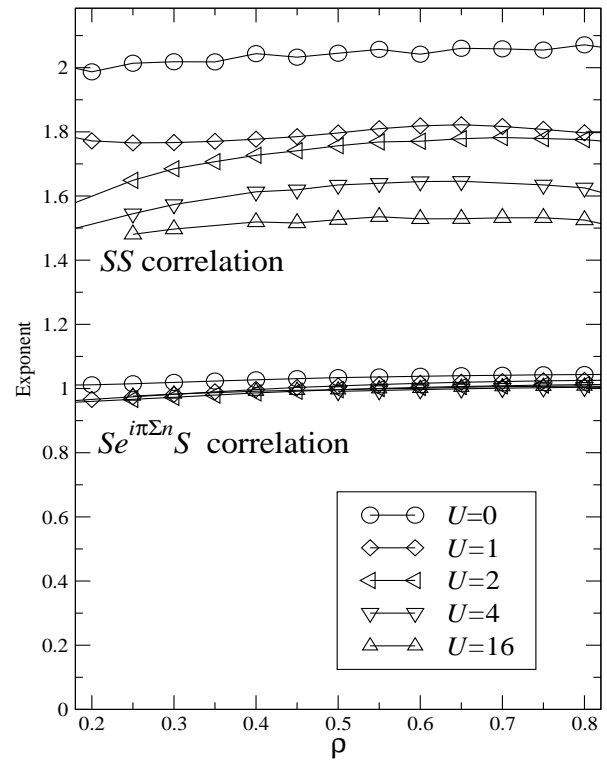}
\end{center}
\caption{Exponents of the spin-spin correlation, $\eta$, (top) and
the topological correlation, $\eta_{top}$, (bottom) as a function of
electron filling, for various values of interaction $U$.}
\label{ianpict}
\end{figure}

In fact, this problem is to a certain extent repairable. The operator
entering the string correlator is not $\sigma^3$ itself, but instead
$\Pi_l \sigma^3 = \exp ( i\pi \int_{x}^{x'} dx \: n(x))$, taking the continuum 
limit. The integral smooths out the discreteness, and an integral over
the continuous bosonization fields can be manipulated to mimic the 
product of Ising valued operators. This is a subtle UV regularization issue,
which we will discuss elsewhere in detail. To give the reader some feeling,
let us consider the expectation value of the string correlator itself,
$D (x) = \langle (-1)^{\sum_{j=0}^{x} n_j} \rangle = \langle 
\cos \left[ \pi \sum_j n (j) \right] \rangle \rightarrow \langle         
\cos \left[ \pi \int_{0}^{x} dy ~ n(y) \right] \rangle$. 
According to bosonization, the total charge density is 
$n (x) = \sqrt{\frac{2}{\pi}}
 \frac{\partial \varphi_c}{\partial x}(j) + {\cal O}_{CDW}(j)
 + {\cal O^\dagger}_{CDW}(j)$ where ${\cal O}_{CDW}(x) = \exp (-2ik_F x) 
\exp (i \sqrt{2 \pi} \varphi_c )  \cos(\sqrt{2 \pi} \varphi_s)$.
It follows that $D (x) \sim  \langle \exp \left(i \sqrt{2 \pi} 
\left[ \varphi_c(x) -\varphi_c(0)  \right] \right)  \rangle
=  \cos(2k_F x) / x^{K_\rho}$. On the other hand, because
on the lattice $(-1)^{n_i} \equiv (-1)^{2S^z_i}$ we might as well
calculate $D (x) = \langle \cos\left[2 \pi \sum_{j=0}^{x} S^z(j) \right]
\rangle$ and using that in bosonization $2S^z (x) = 
\sqrt{\frac{2}{\pi}}
\frac{\partial \varphi_s}{\partial x}(j) 
+ {\cal O}_{SDW}(j) +  {\cal O}^\dagger_{SDW}(j)$ it follows that
$D(x) = \frac{1}{x^{K_\sigma}}$. For $U=0$, $K_\sigma= K_\rho$ and the two 
expressions are the same, but away from this point  
$K_\sigma \neq K_\rho$ and bosonizations runs into a paradox: depending the
way one calculates $D$ one obtains different, and mutually exclusive 
answers. In fact, the DMRG result shows that
$D(x) = A / x^{K_\sigma} + B \cos (2k_F x) / x ^{K_\rho}$ ($A, B$
are non-universal amplitudes), and apparently the two ways of calculating
$D(x)$ in bosonization recover part of the correct answer. 
A similar  problem occurs when one uses bosonization to calculate the
expectation value of $\Pi_l (-1)^{2 n_l}$ which is on the lattice 
just the identity operator; according to bosonization this decays
like $ 1 \over x^{4K_\rho}$! As we will discuss elsewhere\cite{bosonization}, 
these problems can be traced back to spurious contributions associated with the
`spreading' of the electron number density.

In summary, we have
discovered a correlation function which makes it possible to measure
directly the presence of a hidden order underlying spin-charge
separation in 1+1D electron systems. This hidden order can 
be viewed as either a form of topological order, or as
a geometrical structure where the spin system lives in
a squeezed space different from the full chain. The suspicion
was widespread that the Ogata-Shiba construction was special to
the $U \rightarrow \infty$ limit and our main result is the 
numerical demonstration that squeezed space is generic in the
scaling limit. We perceive this as a deep insight, powerful enough
to hint at the existence of hitherto unidentified states of 1+1D
matter. An exciting possibility is closely related to recent 
ideas regarding a potential connection between Ising gauge theory
and the destruction of stripe order in 2+1 dimensions\cite{philmag,sachdev}. As
we discussed, $(-1)^{n_i}$ is Ising valued and the string operator
can be written accordingly as $\Pi_i \sigma^3_i$: this is 
just the Wilson line of $Z_2$ gauge theory\cite{Kogut}. Hence, if local Ising
symmetry would be is dictating, $\langle S \Pi \sigma^3 S \rangle
\sim \exp( x / \xi_g) \langle S S \rangle$ because the l.h.s. is
gauge invariant while the r.h.s. is not; $\xi_g$ is the length scale
where the gauge invariance emerges. In the Luttinger liquid the two
are related by an algebraic factor $ 1 / x^{K_{\rho}}$ but this is
due to the algebraic order in the charge system. Therefore, under the
conditions that (a) the spin system stays separated from the charge,
(b) the charge quanta are bound to sublattice parity flips,
and (c) the charge correlations become short range, one will obtain
the required relations between the correlation functions signaling
the $Z_2$ gauge invariance. Condition (c) is generally satisfied when
the superconducting phase is forced to condense by applying an external
Josephson field. The attractive $U$ Hubbard model fails in this regard
because its ground state can be viewed as a continuation of the local
pair limit where the spin system is destroyed. However, it 
appears\cite{ladders} that
the conditions might well be satisfied in $t-J$ ladders under the influence
of an external Josephson field, and these have in turn much more
similarity with the situation in 2+1D.   

{\em Acknowledgments.} 
We acknowledge stimulating discussions with S.A. Kivelson,
S. Sachdev, F. Wilczek, T. Becker, E. Fradkin, 
N. Nagaosa and T.K. Ng. Financial support was provided by the Foundation
of Fundamental Research on Matter (FOM), which is sponsored by the
Netherlands Organization for the Advancement of Pure Research (NWO). 
Numerical calculations were performed at the National Facility
of the Australian Partnership for Advanced Computation, through the
Australian National University Supercomputer Time Allocation Committee.

\end{multicols}

\end{document}